\documentclass[aps,pra,showkeys,twocolumn]{revtex4}

\usepackage{graphicx}

\begin{document}

\title{On the resolution of quantum paradoxes\\
by weak measurements}

\author{Holger F. Hofmann}
\email{hofmann@hiroshima-u.ac.jp}
\affiliation{
Graduate School of Advanced Sciences of Matter, Hiroshima University,
Kagamiyama 1-3-1, Higashi Hiroshima 739-8530, Japan}
\affiliation{JST,CREST, Sanbancho 5, Chiyoda-ku, Tokyo 102-0075, Japan
}

\begin{abstract}
In this presentation, I argue that weak measurements empirically support the notion 
of quantum superpositions as statistical alternatives. In short, weak measurements
show that Schr\"odinger's cat is already dead or alive before the measurement. 
The collapse of the wavefunction in a strong measurement should therefore be separated 
into the statistical selection of one of the available alternatives and a physical 
interaction that causes decoherence. The application to entanglement reveals that 
measurements in A have no physical effect in B, resolving the paradox of Bell`s 
inequality violation in favor of locality and against (non-empirical) realism. 
\end{abstract}


\maketitle

\section{Introduction}

As Schr\"odinger famously illustrated in his paradox of a cat in a superposition of
life and death \cite{Sch35}, 
it is not at all obvious how one can reconcile the intuitive notion
of an external reality with the concept of quantum coherence between distinguishable
alternatives. Although decoherence may seem to solve the problem, Bell correctly
pointed out that the theory would have to clarify at what point the quantum ``and''
turns into a classical ``or'' \cite{Bel90}. Bohr would probably have objected to this criticism
by pointing out that only the results of experiments are real, and that it is the purpose
of the theory to describe this reality, not to define it. In fact, our only hope for
a scientific resolution of the quantum measurement problem is the design of
experiments that provide empirical answers to our questions. 

In 1988, Aharonov, Albert, and Vaidman proposed such an empirical approach in the
form of weak measurements \cite{Aha88}. Essentially, weak measurements can access the
quantum statistics of a system between its initial preparation $i$ and a final
measurement $f$ with negligible measurement interaction. Recently, this method 
has been applied to experimental realizations of quantum paradoxes, suggesting 
that an empirical resolution of the
paradoxes can be achieved by interpreting the weak values observed in the measurements
as negative probabilities \cite{Wil08,Gog09,Lun09,Yok09}. This seems to be an exciting possibility, but it 
assumes that the negative probabilities observed in weak measurements are well 
understood and not at all paradoxical. Otherwise, we have merely replaced one
paradox with another. 

In the following, I will argue that the real resolution of quantum paradoxes by
weak measurement should be based on the complete quantum statistics of the
states between preparation $i$ and measurement $f$ that can be experimentally
determined by weak measurement tomography \cite{Hof09}. Specifically, weak measurement 
tomography shows that the only empirically correct answer to Bell's question is 
that quantum superpositions are actually statistical alternatives. The quantum 
mechanical ``and'' is a misinterpretation, it is always an ``or''. 
Once this misunderstanding is cleared up, it is possible to resolve 
quantum paradoxes by identifying the fallacies that cause the contradictions
with reality.

\section{Weak measurement tomography}

The key to understanding weak measurements is quantum statistics. An individual
weak measurement results in a completely random outcome, with no discernible
relation to the properties of the quantum state. However, the average of a 
large number of weak measurements reveals small differences in the probability
distribution that can be identified with the quantum statistics of the input state.
In terms of measurement theory, the operator defining the probability of obtaining 
an outcome $m$ in any given measurement can be written as
\begin{equation}
\hat{E}_m = w_m (\hat{1}+\epsilon \hat{S}_m),
\end{equation}
where $||\epsilon \hat{S}_m|| \leq 1$. Thus, the small differences between the experimentally
observed probabilities $p(m)=\mbox{Tr}\{\hat{\rho}_i \hat{E}_m\}$ and $w_m$ can be used
to determine the expectation values of $\hat{S}_m$. The complete density matrix $\hat{\rho}_i$
can be reconstructed from the expectation values of at least $d^2$ linearly independent 
operators $\hat{S}_m$. However, weak measurements have negligible back-action, so the
quantum state is still available for further measurements. It is therefore possible to
determine the quantum statistics of a system between the initial preparation $i$ 
given by $\hat{\rho}_i$ and a final measurement outcome $f$ given by a measurement 
operator $\hat{\Pi}_f$. As explained in more detail in \cite{Hof09}, the joint probabilities 
of the measurement outcomes $m$ and $f$ can be given as 
\begin{equation}
p(m,f|i)=\mbox{Tr}\{\hat{E}_m \frac{1}{2}\left(\hat{\rho}_i \hat{\Pi}_f 
+ \hat{\Pi}_f \hat{\rho}_i \right)\}.
\end{equation}
The measurement $f$ therefore defines a decomposition of the initial density
matrix $\hat{\rho}_i$ into a mixture of sub-ensemble density matrices $\hat{R}_{if}$
associated with the individual outcomes $f$,
\begin{equation}
\label{eq:decomp}
\hat{\rho}_i = \sum_f p(f|i) \; \hat{R}_{if},
\end{equation}
where $p(f|i)=\mbox{Tr}\{\hat{\rho}_i \hat{\Pi}_f \}$ and 
\begin{equation}
\hat{R}_{if} = \frac{1}{2 \mbox{Tr}\{\hat{\rho}_i \hat{\Pi}_f \}}
\left(\hat{\rho}_i \hat{\Pi}_f 
+ \hat{\Pi}_f \hat{\rho}_i \right).
\end{equation}
This decomposition is valid even if the density matrix $\hat{\rho}_i$
describes a pure state that is a coherent superposition of different outcomes $f$.
In other words, eq. (\ref{eq:decomp}) shows that there is no contradiction 
between a statistical interpretation of the alternatives $f_1$ and $f_2$ and
quantum coherence between the eigenstates associated  with those outcomes.
Specifically, the problem is resolved by attributing half of the coherence to
$f_1$ and the other half to $f_2$. 

\section{Schr\"odinger's cat and double slit interference}

The significance of the result derived above emerges when it is applied to 
the examples that are used to illustrate the strangeness of quantum mechanics.
Firstly, eq.(\ref{eq:decomp}) suggests that the death or survival of
Schr\"o\-dinger`s cat is not decided by the measurement - the measurement merely
uncovers which of the alternatives has already happened. 

It may be surprising that a simple argument about quantum measurements should 
lead to new insights into the nature of quantum coherence. However, it seems
that much of the perplexity caused by Schr\"o\-dinger's cat was due to the
analogy of quantum states and classical waves, where interference patterns 
appear to require the presence of two sources. It is therefore useful to 
consider the statistical explanation of coherence in terms of the double
slit interference of a single particle. Here, it is commonly argued that
the interference pattern could not exist if the particle either went through
one slit or through the other slit. However, weak measurement tomography 
shows that the states defined by a two slit superposition 
$\mid 1 \rangle + \mid 2 \rangle$ and a subsequent which-path measurement are
\begin{eqnarray}
\hat{R}_{+1} &=& \mid 1 \rangle \langle 1 \mid +
\frac{1}{2}\left(\mid 1 \rangle \langle 2 \mid
+ \mid 2 \rangle \langle 1 \mid \right)
\nonumber \\
\hat{R}_{+2} &=& \mid 2 \rangle \langle 2 \mid +
\frac{1}{2}\left(\mid 1 \rangle \langle 2 \mid
+ \mid 2 \rangle \langle 1 \mid \right).
\end{eqnarray}
Thus, weak measurement shows that the assumption that which-slit information
prevents interference is unfounded - the interference pattern
associated with the coherences between path 1 and path 2 shows up equally in
the weak measurements performed on photons found in path 1 and on photons 
found in path 2. 

By analogy, the same should be true for Schr\"o\-dinger's cat. Theoretically, 
there exists a microscopic interference pattern that arises from the
superposition of death and survival. If we were to perform weak measurement
tomography on the cat before confirming its death or survival, we would find that
(a) the dead cats were already dead when the tomography was performed and
(b) both dead and living cats showed the microscopic signature of a ``+''
superposition before the final measurement confirmed their death or survival.

\section{Entanglement and non-locality}

Entanglement is widely considered to be the most counter-intuitive feature
of quantum mechanics. In many introductions of entanglement, it is assumed that
the non-local change of the wavefunction caused by a local measurement
is an effect that has no classical explanation. However, such claims are rather
misleading, since the same non-local change is caused by a classical (Bayesian)
update of probabilities if the initial probability distribution (or prior) 
includes correlations between the two systems. In the following, I will show that
this is indeed the correct analogy. 

Consider a maximally entangled state $\mid E \rangle$ of two $d$-level systems, 
$A$ and $B$. The notion of non-locality arises because any measurement represented by
a projection on a pure state $\mid f \rangle$ in $A$ results in a corresponding
pure state $\mid f^* \rangle$ in $B$,
\begin{equation}
_A \langle f \mid E \rangle_{AB} = \frac{1}{\sqrt{d}} \mid f^* \rangle_B.
\end{equation}
However, the density matrix of the maximally entangled state can already be
written as a mixture of $d$ alternatives $f$ given by
\begin{eqnarray}
\hat{R}_{Ef} &=& \frac{d}{2}\big((\mid E \rangle \langle E \mid)_{AB}
(\mid f \rangle \langle f \mid)_A 
\nonumber \\ && \hspace{0.5cm}
+(\mid f \rangle \langle f \mid)_A (\mid E \rangle \langle E \mid)_{AB}\big)
\nonumber \\
&=& \frac{\sqrt{d}}{2} \left(\mid E \rangle \langle f; f^* \mid 
+ \mid f; f^* \rangle \langle E \mid \right).
\end{eqnarray}
Here, the coherences between $\mid f; f^* \rangle$ and orthogonal components of
$\mid E \rangle$ represent the non-local correlations between properties of 
$A$ and $B$ other than $f$ and $f^*$. Weak measurement tomography therefore
shows that (a) the property $f^*$ was already present in $B$ before the measurement
in $A$, and (b) the two systems are strongly correlated in the properties of
the systems other than those determined by $f$ and $f^*$ until the correlations
are lost in the back-action related randomization of the local properties 
in system $A$. 

One significant consequence of this analysis is that the physical back-action of
the measurement is completely local. The changes of the quantum state in $B$ 
are completely explained by the selection of an alternative already present in 
the initial density matrix. To confirm that system $B$ is already in the pure
state $\mid f^* \rangle$ even before the measurement in $A$ is performed, we 
can trace out system $A$
 to find the local statistics described by $\hat{R}_{Ef}$,
\begin{equation}
\mbox{Tr}_{A}\{ \hat{R}_{Ef}\} =
d \langle f \mid E \rangle \langle E \mid f \rangle = 
\mid f^* \rangle \langle f^* \mid. 
\end{equation}
Initially, system $B$ is in an incoherent mixture of the states $\mid f^* \rangle$,
and the measurement in $A$ merely identifies the alternative that is actually
realized in a given individual case. Thus, the non-locality of entanglement is
not really different from the non-locality of classical correlated statistics.
The difference between classical physics and quantum physics lies elsewhere.

\section{Negative probability paradoxes}

All quantitative paradoxes in quantum mechanics are based on the simultaneous
assignment of values to measurement results that cannot be obtained at the
same time. Weak measurements show that the joint probabilities associated with
such simultaneous assignments of eigenvalues can be negative. In weak measurement
tomography, the possibility of negative joint probabilities shows up in the
form of negative eigenvalues of the density matrix $\hat{R}_{if}$. 

If $\hat{\Phi}_g$ is the operator for the probability of obtaining an outcome 
$g$ in a given measurement, then the probability of $g$ obtained in weak measurements
is given by
\begin{equation}
p(g|i,f) = \mbox{Tr} \{\hat{R}_{if} \hat{\Phi}_g \},
\end{equation}
as in conventional measurement theory. This formalism accurately describes the
negative probabilities observed in experimental resolutions of quantum paradoxes 
by weak measurements such as the one reported in \cite{Yok09}. Alternatively, it is possible
to define the joint probability of $f$ and $g$ for the initial state $\hat{\rho}_i$,
as shown in \cite{Hof09}. The result is a joint probability given by
\begin{equation}
p(f,g|i) = \mbox{Tr} \{\hat{\rho}_i \frac{1}{2} \left(
\hat{\Pi}_f \hat{\Phi}_g + \hat{\Phi}_g \hat{\Pi}_f
\right)\}.
\end{equation}
Since this definition of joint probabilities is the only one consistent with
weak measurement tomography, it is independent of the specific experimental
realization which is used to confirm it. In particular, it does not matter
whether $g$ or $f$ is obtained in the weak measurement. Moreover, the joint
probability of $g$ and $f$ is consistent with all strong measurements of 
$g$ and $f$. It is therefore possible to explain quantum paradoxes constructed
from the measurement results of strong measurements in terms of the negative
joint probabilities observed in weak measurements.

Since it was shown in the previous section that entanglement does not involve any
physical non-locality, it may be of particular interest to explain
Bell's inequality violation in terms of negative joint probabilities.
In the formulation of Bell's inequalities, values of $\pm 1$ are assigned to
orthogonal components $\hat{X}$ and $\hat{Y}$ of a two level system.
The operators $\hat{\Pi}(X,Y)$ describing the joint probabilities can be obtained from 
the projectors $(1\pm \hat{X})/2$ and $(1\pm \hat{Y})/2$. They read
\begin{eqnarray}
\hat{\Pi}(+1;+1) &=& \frac{1}{4}(1+\hat{X}+\hat{Y})
\nonumber \\
\hat{\Pi}(+1;-1) &=& \frac{1}{4}(1+\hat{X}-\hat{Y})
\nonumber \\
\hat{\Pi}(-1;+1) &=& \frac{1}{4}(1-\hat{X}+\hat{Y})
\nonumber \\
\hat{\Pi}(-1;-1) &=& \frac{1}{4}(1-\hat{X}-\hat{Y}).
\end{eqnarray}
The eigenvalues of each of these operators are $(1 \pm \sqrt{2})/4$, corresponding
to probabilities of 60 \% and -10 \%. The eigenstates are those of the spin operator 
$\hat{S}_+$ along the diagonal between $\hat{X}$ and $\hat{Y}$ for 
$\hat{\Pi}(+1;+1)$ and $\hat{\Pi}(-1;-1)$, and those of the 
orthogonal spin component $\hat{S}_-$ for $\hat{\Pi}(+1;-1)$ and $\hat{\Pi}(-1;+1)$.
If the initial state is maximally entangled, the values of $S_+=\pm 1$ and $S_-=\pm 1$
in system $B$ can be determined by measurements in system $A$. The joint probabilities
for $X$, $Y$, $S_+$ and $S_-$ then result in the following total probabilities,
\begin{center}
\begin{tabular}{rll}
60 \% & for & $(X+Y) S_+ = +2$ 
\\
& and & $(X-Y) S_- = 0$
\\
60 \% & for & $(X+Y) S_+ = 0$ 
\\
& and & $(X-Y) S_- = +2$
\\
-10 \% & for & $(X+Y) S_+ = -2$ 
\\
& and & $(X-Y) S_- = 0$
\\
-10 \% & for & $(X+Y) S_+ = 0$ 
\\
& and & $(X-Y) S_- = -2$.
\end{tabular}
\end{center} 
Clearly, the value of $(X+Y)S_+ + (X-Y)S_-$ is either $+2$ or $-2$. However,
local negative probabilities result in an average of $2 \sqrt{2}$. 

This result can be confirmed experimentally by weak measurements of the
joint probabilities. Significantly, the negative probabilities originate not from
some mysterious action at a distance, but from the local correlations between
the spin components of the same system. The paradox of Bell's inequality violation
is therefore resolved in favor of locality and against realism.

\section{The problem with reality}

Weak measurement tomography shows that a measurement result $f$ can be considered
real even before the measurement is performed. Specifically, the quantum statistics
of a pure state $\hat{\rho}_i=\mid \psi \rangle \langle \psi \mid$ can be decomposed 
into a mixture of non-positive statistical operators $\hat{R}_{if}$ in anticipation 
of the measurement $f$. However, the decomposition is only justified if the 
measurement of $f$ is actually performed. The negative joint probabilities of 
measurements that cannot be performed jointly indicate that reality cannot be
simultaneously attributed to the outcomes $f$ and $g$ of both measurements.

Weak measurements therefore indicate that quantum paradoxes arise from the assumption
of a measurement independent reality. In everyday life, we naturally assume that
such a reality exists. Obviously, it would be crazy to assume that a room ceases
to exist when we close the door from the outside. However, it may be foolish to 
extrapolate this experience without reflecting on its foundations. In fact, almost
every philosopher who thought deeply about the origin of our sense of reality
noticed that it is somehow rooted in experience and perception. The list includes 
famous names such as Descartes (``I think therefore I am''), Berkeley (``to exist is 
to be perceived''),  Kant (no experience of the ``thing in itself''), and 
Schopenhauer (no object 
without subject). In practice, objective reality is established through the 
consistency of observations. As Kant explained in perhaps excessive detail, we 
identify realities by extending our experience using chains of causality relations. 
Normally, this results in a tightly knit web with very little wiggle space -
especially in the physical sciences. Nevertheless, the primary condition for the
reality of an object is always the possibility of experiencing its effects.

The uncertainty principle of quantum mechanics defines an absolute limit for
the effects of an individual quantum object on the web of empirical reality.
However, quantum mechanics can provide a complete description of the 
measurement statistics obtained from a certain class of objects. 
Intuitively, we try to identify the individual representative with the whole
class, and are therefore confused by the appearance of irreconcilable 
differences between representatives observed in different ways. However, there
is really no reason why a particle with the observable property $f$ should
also have a hidden property $g$, just because this property  has been observed on
a different member of the same class of objects.

Ultimately, our imagination is pushed to its limits when it comes to quantum
mechanics. Nevertheless, we might achieve some insights by identifying the correct
analogies and rejecting false and misleading ones. As far as reality is concerned,
it may be useful to discard the notion of fundamental material objects in favor of
a model that emphasizes the role of perspective. Quantum objects seem to show us
only their front, whereas the backside is hidden by uncertainty. When we look
at different objects of the same class, we can construct something that looks like
a complete image, but there seem to be inconsistencies when we try to identify
the backside of an individual object. It is almost as if the individual quantum
objects were stage props showing different sides of the same building in
different scenes of a play. In our minds, we construct the complete building,
and for the purpose of understanding the action on the stage, this is quite useful.
But a look backstage would show that the people who built the set did not really
bother to construct the whole building, but achieved their purpose much more
economically with just a bit of wood and canvas. Now, should we be disappointed
by this absence of ``realism''? If the world is a stage, this might actually depend on 
whether the kind of realism we have in mind has any impact on our appreciation of
the plot.

\section{Conclusions}

Empirically, quantum mechanics is a statistical theory that predicts only the
probabilities of events. In the case of weak measurements, it is particularly
important to keep this in mind since the results of weak measurements are
averages over a large number of individual measurements. Nevertheless weak
measurements can provide a complete description of a quantum state defined by
both initial conditions $i$ and final conditions $f$. The non-positive density 
matrix $\hat{R}_{if}$ of this state can be determined experimentally by weak 
measurement tomography \cite{Hof09}. 

Weak measurement tomography shows that the collapse of the wavefunction caused by
the final measurement $f$ is very similar to the collapse of a classical 
probability distribution when additional information becomes available. In fact, 
the initial state $\hat{\rho}_i$ can always be written as a mixture of the
non-positive density matrices $\hat{R}_{if}$ associated with the different possible
outcomes $f$. Coherences between different $f$ are divided equally between the
alternative outcomes. 

In the case of entangled states, the non-locality of the collapse of the wavefunction
can
 be explained entirely in terms of the statistical correlations between the 
systems. The non-locality is therefore statistical and corresponds to the non-locality
of an update of probability in classically correlated systems. Quantum paradoxes 
arise only from the negative eigenvalues of $\hat{R}_{if}$ that define negative
joint probabilities for the outcome $f$ and the outcome $g$ of an alternative 
measurement. 

While it is possible to attribute reality to the measurement result $f$ even before
the measurement happens, it is not possible to attribute reality to a measurement
result that is not actually obtained. The negativity of joint probabilities shows 
that any assignment of hidden variables would be inconsistent with the empirical 
results obtained in weak measurements. Weak measurements therefore indicate that 
non-empirical realism (that is, the dogmatic insistence on a measurement independent
reality) is inconsistent with the statistical predictions of quantum mechanics.
Thus, weak measurements resolve quantum paradoxes by identifying non-empirical
realism as the crucial fallacy in the formulation of the paradox.

\section*{Acknowledgements}
Part of this work has been supported by the Grant-in-Aid program of the
Japanese Society for the Promotion of Science, JSPS


\vspace{0.5cm}

\begin{thebibliography}{xyz00}


\bibitem{Sch35}
E. Schr\"odinger, Naturwissenschaften {\bf 23}, 807; 823; 844 (1935).

\bibitem{Bel90}
J.S. Bell, Physics World {\bf 3}, issue 8, 33 (1990).

\bibitem{Aha88}
Y. Aharonov, D. Z. Albert, and L. Vaidman, Phys. Rev. Lett. {\bf 60}, 1351 (1988).


\bibitem{Wil08}
N. S. Williams and A. N. Jordan, 
Phys. Rev. Lett. {\bf 100}, 026804 (2008).

\bibitem{Gog09}
M. E. Goggin, M. P. Almeida, M. Barbieri, B. P. Lanyon, J. L. O'Brien, A. G. White, and
G. J. Pryde, e-print arXiv: 0907.1679v1.

\bibitem{Lun09}
J. S. Lundeen and A. M. Steinberg, Phys. Rev. Lett. {\bf 102}, 020404 (2009).

\bibitem{Yok09}
K. Yokota, T. Yamamoto, M. Koashi, and N. Imoto, New J. Phys. {\bf 11},
033011 (2009)


\bibitem{Hof09}
H.F. Hofmann, arXiv:0907.0533v2

\end{thebibliography}
\end{document}